\documentclass[pdflatex,sn-mathphys-num]{sn-jnl}

\usepackage{graphicx}%
\usepackage{multirow}%
\usepackage{amsmath,amssymb,amsfonts}%
\usepackage{amsthm}%
\usepackage{mathrsfs}%
\usepackage[title]{appendix}%
\usepackage{xcolor}%
\usepackage{textcomp}%
\usepackage{manyfoot}%
\usepackage{booktabs}%
\usepackage{algorithm}%
\usepackage{algorithmicx}%
\usepackage{algpseudocode}%
\usepackage{listings}%
\usepackage{enumitem}%
\usepackage{array, booktabs, multirow, tabularx}%

\theoremstyle{thmstyleone}%
\theoremstyle{thmstyletwo}%

\theoremstyle{thmstylethree}%

\begin{document}

\title[Augmented Question-guided Retrieval (AQgR) of Indian Case Law with LLM, RAG, and Structured Summaries
]{Augmented Question-guided Retrieval (AQgR) of Indian Case Law with LLM, RAG, and Structured Summaries
}


\author*[1]{\fnm{Vishnuprabha} \sur{V}}\email{vishnuprabha72@gmail.com}

\author[1]{\fnm{Daleesha} \sur{M Viswanathan}}
\email{daleesha@cusat.ac.in}
\author[2]{\fnm{Rajesh} \sur{R}}

\author[3]{\fnm{Aneesh} \sur{V. Pillai}}\email{dr.avpillai@cusat.ac.in }

\affil*[1]{\orgdiv{Division of Information Technology, School of Engineering}, 
\orgname{Cochin University of Science and Technology}, 
\country{India}}

\affil[2]{
\orgname{NPOL}, 
\country{India}}

\affil[3]{
\orgdiv{School of Legal Studies}, 
\orgname{Cochin University of Science and Technology}, 
\country{India}}


\abstract{Finding relevant legal precedents for specific situations can be challenging for lawyers. Most current methods focus on finding similar cases based on the facts rather than analyzing the underlying legal issues. Additionally, existing systems for retrieving case law often fail to provide explanations on how a particular case is relevant to the given facts. This paper proposes the utilization of Large Language Models (LLMs) to address this issue, despite their limitations in handling lengthy texts. The proposed approach combines Retrieval Augmented Generation (RAG) with structured summaries tailored for Indian case law retrieval to overcome context length constraints. By generating legal questions based on the factual scenario using AQgR (Augmented Question guided Retrieval) approach, the system aims to identify relevant case laws. Experimental testing conducted on a small subset of the FIRE 2019\cite{Bhattacharya2019} dataset demonstrated promising results, achieving a Mean Average Precision score of 0.36 and a Mean Average Recall of 0.67 across a set of queries, exceeding the current MAP value of 0.1573\cite{LeburuDingalo2020}. The proposed method not only retrieves relevant case law but also provides explanations on the relevance of these precedents to the current factual scenario. The adoption of legal questions as part of the case law retrieval process could prove beneficial for legal professionals, particularly judges. Furthermore, the structured summaries generated by this approach hold potential for various legal tasks, such as statute retrieval, judgment prediction, and argument generation.}


 \abstract{\textbf{Context}: Identifying relevant legal precedents remains challenging, as most retrieval methods emphasize factual similarity over legal issues, and current systems often lack explanations clarifying case relevance. \textbf{Objective}: This paper proposes the use of Large Language Models (LLMs) to address this gap by facilitating the retrieval of relevant cases, generating explanations to elucidate relevance, and identifying core legal issues—all autonomously, without requiring legal expertise. The introduction of the structured summary approach for case laws helps to addresses common LLM limitations associated with lengthy legal texts. \textbf{Method}: Our approach combines Retrieval Augmented Generation (RAG) with structured summaries optimized for Indian case law.. Leveraging the Augmented Question-guided Retrieval (AQgR) framework, the system generates targeted legal questions based on factual scenarios to identify relevant case law more effectively. \textbf{Results}: The structured summaries were assessed manually by legal experts, given the absence of a suitable structured summary dataset. Case law retrieval was evaluated using the FIRE dataset, and explanations were reviewed by legal experts, as explanation generation alongside case retrieval is an emerging innovation. Experimental evaluation on a subset of the FIRE 2019 dataset yielded promising outcomes, achieving a Mean Average Precision (MAP) score of 0.36 and a Mean Average Recall (MAR) of 0.67 across test queries, significantly surpassing the current MAP benchmark of 0.1573. \textbf{Conclusion}: This work introduces a suite of novel contributions to advance case law retrieval. By transitioning from fact-based to legal-issue-based retrieval, the proposed approach delivers more contextually relevant results that align closely with legal professionals' needs. The addition of explanation generation alongside case retrieval offers essential contextual insight, enabling users to assess case applicability more accurately. Tailored structured summaries for Indian case law provide concise, jurisdiction-specific representations that support a range of legal applications. Moreover, integrating legal questions within the retrieval process through the AQgR framework ensures more precise and meaningful retrieval by refining the context of queries. These innovations constitute a comprehensive solution with significant potential for diverse legal tasks, including statute retrieval, judgment prediction, and argument generation.}




\maketitle

\section{Introduction}
Legal professionals worldwide, particularly in nations with common law or mixed legal systems, intensively engage in case law retrieval as a core element of their practice. India operates under a hybrid legal framework combining common law principles and written statutes to settle legal conflicts. The common law rooted in former judicial decisions(precedents) differs from the statute or written laws, including rules and regulations protecting a citizen's fundamental rights. 

Locating relevant precedents greatly impacts a lawyer's success in a case \cite{winningcourtcases}. An experienced lawyer who has participated in many trials knows how a judge will frame the issues based on the facts. By knowing the possible issues that a Judge can frame, the lawyer uses his experience and historical knowledge to locate principles(from precedents and statutes) that can be applied to that legal issue to get remedies for the client. To locate the legal principles, a lawyer needs both knowledge and time. The lawyers are charged based on their time on a case to find relevant principles applicable to the facts. Hence, there is a pressing requirement for an automated case law retrieval system to assist legal professionals. 

Automated case law retrieval systems hold great promise for helping young legal professionals build stronger arguments by pinpointing relevant precedents \cite{Ashley1992}. Judges may similarly leverage this tool to reinforce their opinions that rely on precedents and render timely judgments. Streamlined decision-making processes decrease delays and thereby ensure fast access to legal justice.

In India, the huge volume of unstructured judgments, absence of headings or separation between rhetorical roles such as facts, issues, etc., makes the automation process challenging. The existing approaches for Indian precedent retrieval solely focus on ranking a judgment's relevance to a given fact based on a similarity score. A lawyer must read the entire retrieved judgment to know the legal principle applicable to the given fact. 

\begin{figure}[h]
\centering
\includegraphics[width=1.0\textwidth]{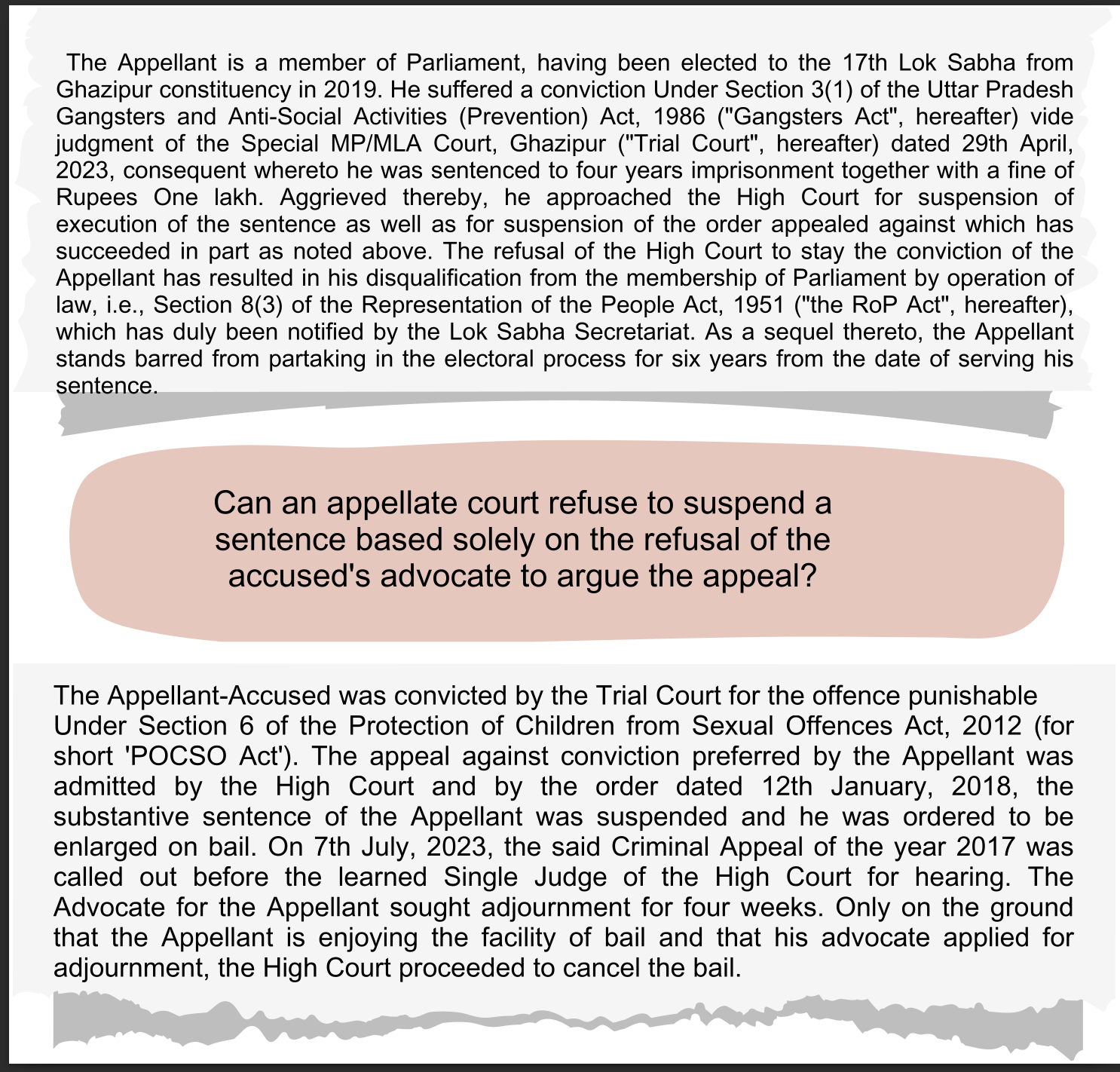}
\caption{Figure shows the example of a legal issue with two different fact scenarios}\label{fig1}
\end{figure}

Another drawback of the current systems is that they rely on facts for case law retrieval. However, in real life, lawyers use legal issues to identify the case laws and not the facts. There will be situations where there will be entirely different fact sets that address the same legal issue. The figure \ref{fig1} shows an example of a legal issue with entirely different facts. So, the similarity-based case law retrieval using facts identifies the judgments with similar facts and not similar legal issues. However, finding legal issues, given the fact, is very difficult using the existing deep learning frameworks. The capacity to identify legal issues comes with historical knowledge and intelligence. However, the generative models can generate legal questions about a given fact as they incorporate historical knowledge and intelligence.

In short, the existing automated system for retrieving precedents in Indian judgments falls short of conveying a precedent's relevance or how it applies to a given fact. This research aims to address this gap with the potential abilities of the Large Language Model(LLM) \cite{radford2018improving} to generate text by understanding the context. An automated case law retrieval with application principle can save time and effort spent by lawyers to a great extent.

Despite the potential ability to understand and generate content, LLM has a few limitations, too\cite{bubeck2023sparks}. \textit{Hallucination:} LLMs might imply the existence of non-existent case laws. \textit{Context Length Limitations:}Many Indian legal judgments span extensively, presenting challenges for LLMs with smaller context windows. Efficient precedent retrieval necessitates maximizing the number of accommodated precedents within an LLM's context window. \textit{Limited Control over the LLM :} Controlling LLM output remains extremely challenging. An LLM can generate multiple results for multiple runs. \textit{Cost:} Usage expenses for LLMs vary depending on token counts and context window sizes. Accommodating more cases to context windows incurs higher costs.\textit{Outdated knowledge:} LLM needs to be retrained to update its knowledge regularly to utilize it for high-end applications.

Regardless of these limitations,  this research focuses on LLMs' practical utility and effectiveness in case law retrieval with explanation, expecting reduced usage cost and increased context length in future. This research also considers the value of structured or template-based summaries in case law retrieval.A template-like summary allows an advanced search to identify similar cases based on facts, statutes, precedents, questions of law, and other variables. However, the efforts to generate such a template-based summary are almost zero for Indian legal Judgments. Generating a structured summary with deep learning requires different models to extract different legal concepts. However, an LLM trained to solve multiple tasks is well suited for automatic template-based summary generation of Indian legal judgments. Creating structured summaries for case law retrieval with legal experts is time-consuming and expensive, especially for long judgments. The structured summaries produced by LLMs not only help laypeople get a clear idea about cases but also help solve the context length limitations of LLMs. Retrieval Augmented Generation (or RAG)\cite{Lewis2020} is a framework that allows the integration of external databases with LLM without the need to fine-tune an LLM for domain-specific data. A Retrieval Augmented Generation approach that retrieves relevant content that fits the LLM context window from an external database is used for structured summary generation and case law retrieval. RAG offers a solution for handling context-length issues in LLM.

Accordingly, this research addresses the following questions:
\begin{itemize}
\item How do LLMs perform in case law retrieval withstanding context length restrictions and without fine-tuning?
\item How do structured summaries contribute to case law retrieval using LLMs?
\item Contribution of legal issue in case law retrieval rather than facts.
\item Can Retrieval Augmented Generation (RAG) be used with LLMs to manage lengthy legal documents and diminish LLM hallucination?
\item How proficient is RAG in generating structured or template-like summaries with just one query and zero-shot learning?
\end{itemize}

Contributions to the legal and AI communities include:

\begin{itemize}

\item \textbf{Developed a Structured Template and Summary Generation System for Indian Legal Judgments}: This work introduces a structured template specifically designed for Indian legal judgments, as shown in Table \ref{tab:summary}
. An automated system was developed to generate summaries from unstructured judgments by prompt-tuning large language models (LLMs). Using the Gemini-pro model with a single query and zero-shot learning, the system generates structured summaries of Indian legal judgments. This novel approach to automatically generating structured summaries from judgments has significant applications across various legal domains.

\item \textbf{Introduced an innovative approach for case law retrieval using LLMs}: A novel method for retrieving case law was introduced, which leverages large language models without requiring fine-tuning. The approach also incorporates explanations and uses questions of law to improve the retrieval process.

\item \textbf{Improved retrieval efficiency through structured summaries and key legal issue identification:} The research highlights the effectiveness of using structured summaries with predefined fields to enhance the efficiency of case law retrieval. Additionally, it emphasizes that identifying the key legal issue is critical for successful retrieval. The ability of LLMs to learn from context without extensive training was also verified.

\item \textbf{Proposed a scalable method for case law retrieval in the era of generative AI}: A practical, scalable solution for case law retrieval using LLMs was presented, addressing challenges such as handling lengthy legal documents and refining queries in Retrieval-Augmented Generation (RAG) systems. This approach holds significant potential as generative AI continues to evolve.

\end{itemize}

\section{Related Works}

\begin{table}
\centering
\begin{tabularx}{\linewidth}{*{6}{X}}
\toprule
Author Citations & Task & Input & Output & Jurisdiction \\
\midrule
Vuong et al. \cite{Vuong2023} & Case Law Retrieval & Legal Queries & Relevant Case Laws & Foreign \\
\\
Askari and Verberne \cite{Askari2021} & Case Law Retrieval & Full Case Documents & Potentially Relevant Prior Cases & Foreign \\

\\
Nguyen et al.\cite{Nguyen2022} & Statute Law Document Retrieval & Legal Queries & Relevant Statute Law Documents & Foreign \\
\\
Klein et al. \cite{Klein2006} & Case Law Retrieval & Legal Keywords & Relevant Case Laws & Foreign \\
\\
Maxwell and Schafer \cite{Maxwell2008} & Legal Information Retrieval & Legal Queries & Relevant Legal Documents & Foreign \\
\\

DAgostini Bueno et al.\cite{DAgostiniBueno1999} & Intelligent Retrieval of Jurisprudencial Texts & Legal Queries & Relevant Jurisprudencial Texts & Foreign \\
\\
van Opijnen and Santos \cite{vanOpijnen2017} & Legal Information Retrieval(IR) and Entailment & Legal Documents and Queries & Determined Whether a Query Implies Another Statement or Is Irrelevant & Foreign \\
\\
Kim et al. \cite{kim2019statute} & Legal IR and Entailment & Legal Documents and Queries & Determined Whether a Query Implies Another Statement or Is Irrelevant & Foreign \\
\\
Mandal et al. \cite{Mandal2021} & Legal Document Similarity & Legal Document & Similar Legal Documents & Indian \\
\\
Leburu-Dingalo et al.\cite{LeburuDingalo2020} & Precedent and Statute Retrieval & Legal Descriptions of Current Situations & Relevant Statutes and Prior Cases & Indian \\
\\
Bhattacharya et al. \cite{Bhattacharya2019} & Precedent and Statute Retrieval for Indian Legal Domain & Legal Facts & Relevant Precedents and Statutes & Indian \\
\\
Sampath and Durairaj \cite{Sampath2022} & Precedence Retrieval from Legal Documents & Current Case Documents and Legal Domain & Similar Prior Case Documents & Indian \\
\\
Pandian and Joshi \cite{Pandian2022} & Case Law and Statute Retrieval & Legal Queries & Relevant Case Laws and Statutes & Indian \\
\bottomrule
\end{tabularx}
\caption{Relevant Related works on Case Law Retrieval.}
\label{tab1:litreview}
\end{table}

In legal research, case law retrieval and case law entailment tasks have typically been treated as separate endeavours. The COOLIE competition, as discussed by Kim et al. (2023) \cite{Kim2023}, stands out for its focus on four distinct legal tasks: case law retrieval, case law entailment, statute retrieval, and statute entailment. Case law retrieval involves searching and retrieving relevant legal cases or precedents from extensive repositories of legal documents based on specific queries or criteria. On the other hand, case law entailment refers to determining whether a given legal precedent (the source case) entails the outcome of another legal case (known as the hypothesis case). This entailment finds the paragraph of a retrieved case that applies to the provided facts. However, it can be a reasoning paragraph or a legal issue paragraph or fact paragraph.

In the Indian legal context, a similar competition for case law retrieval exists. The FIRE 2019 competition, detailed by Bhattacharya et al. (2019)\cite{Bhattacharya2019}, focused on two primary tasks: case law retrieval and statute retrieval. In the subsequent FIRE 2020 competition (Bhattacharya et al., 2020)\cite{Bhattacharya2020}, an additional task, rhetorical role labelling, was introduced to address the unstructured nature of Indian legal judgments. The top mean average precision (MAP) score for precedent retrieval in FIRE 2020 was 0.1573\cite{LeburuDingalo2020}, while in FIRE 2019, it stood at 0.1492.

A paper by Sampath et al. (2022)\cite{Sampath2022}  worked with Indian case law retrieval reported an impressive MAP score of 0.632. However, it is worth noting that their approach used whole judgments as queries for identifying related judgments, which is different from a fact-based retrieval. While effective for finding similar legal documents, this method may be less practical in real-world scenarios where lawyers typically work with factual statements or legal questions to identify applicable precedents. The work by \cite{Mandal2021} also measures similarity between legal documents.

From the Table \ref{tab1:litreview} it is evident that almost all the works employed either a legal fact or legal judgment as Input query for case law retrieval or entailment. Most of the fact-based case law retrieval works in Indian scenarios, finds relevant case laws by calculating the similarity between embeddings of a given fact with embeddings of judgments present in the pool. The judgments that possess the highest similarity score with the facts are relevant. The existing approaches use deep learning frameworks to get better matching. However, embedding will have a fixed dimension in a deep learning framework. The fact, which consists of around 1000 words, will be converted into an embedding of a specific dimension, and a whole judgment with 20,000 words will also be converted into an embedding of the same dimension. This approach leads to loss of information and incorrect results for case law retrieval. The highly relevant judgments obtained from the case law retrieval process are subsequently utilized to determine entailment. Again for entailment the input will be fact and the relevant Judgment, the output will be paragraphs of the judgment that most entails the given fact.

Moving into the application of Large Language Models (LLMs) in the legal domain, Vats et al. (2023)\cite{vats2023llms} utilized LLMs for statute prediction and judgment prediction in the context of Indian legal judgments. They reported satisfactory performance in statute prediction but encountered challenges with judgment prediction, noting gender and religion biases in the LLM-generated outputs.

Savelka et al. (2023). \cite{Savelka2023} explored using an augmented LLM to explain legal concepts. While the unaugmented system yielded less coherent results, the augmented approach provided clearer explanations for legal concepts.

Furthermore, Yu et al. (2023)\cite{Yu2023} investigated LLMs for the legal entailment task, determining whether retrieved articles entail the provided query. They found that incorporating a legal reasoning prompt, including Issue, Rule, Application, and Conclusion (IRAC), along with the query and articles, improved legal entailment performance.

This study introduces a novel approach by treating the tasks of case law retrieval and case law entailment as interconnected challenges rather than separate ones. We leverage the capabilities of Large Language Models (LLMs) to address these issues collectively. Also, questions of law or issues are used to retrieve relevant documents instead of facts. Additionally, the research tackles the inherent unstructured nature of Indian legal judgments using LLMs, offering a tailored solution to enhance case law retrieval in this specific context. Furthermore, the paper provides a methodology for effectively managing context length limitations and mitigating output variations within LLMs, adding a valuable dimension to the proposed solution.

\section{Methodology}
\subsection{Dataset}
In the domain of Indian law, a dataset was curated for legal precedent retrieval as part of FIRE 2019\cite{Bhattacharya2019}. This dataset encompasses approximately 2800 judgments and 50 queries, representing factual scenarios or legal contexts necessitating relevant precedents. Due to limited access to Large Language Models (LLMs), a subset of 50 judgments was carefully selected from the FIRE dataset that includes civil and criminal cases. Initially, 15 queries and their corresponding relevant judgments were chosen. Following the exclusion of judgments containing inappropriate content for LLMs, such as sexual or violent material, roughly 35 relevant judgments remained. Subsequently, the remaining 15 judgments were randomly selected from the dataset. Structured summaries were then generated for these 50 judgments using an LLM and saved in JSON format, facilitating their utilization in case law retrieval.

Here is the list of selected queries (Q1, Q2, Q3, Q4, Q5, Q31, Q32, Q33, Q34, Q35, Q46, Q47, Q48, Q49, and Q50) along with the corresponding chosen judgments (C1, C2797, C2801, C141, C54, C92, C121, C14, C122, C2799, C59, C21, C72, C9, C49, C164, C170, C2796, C47, C162, C38, C2803, C25, C93, C94, C184, C139, C69, C22, C2798, C2802, C2804, C152, C182, C27, C85, C76, C31, C2805, C155, C171, C65, C147, C2806, C2800, C126, C186, C75, C82, and C79).

However, the selected judgments did not provide sufficient length to evaluate the capabilities of RAG fully. To address this, a few longer judgments were included for testing purposes.

\subsection{Framework Overview}

The proposed method is built on top of a technique called Retrieval Augmented Generation (or RAG)\cite{Lewis2020}, which enables the LLM to adapt to a domain-specific task without any cost compared to fine-tuning. The fundamental components of the RAG framework include:
\textit{Retrieval:} Retrieves relevant information from an external knowledge source. Vector databases are used as external knowledge sources. Vector database stores the contents as chunks with proper indexing. 
\textit{Generation:}  The retrieved local knowledge is incorporated in the LLM context window, and the LLM's Global knowledge is utilized to generate the response by including local knowledge. The user can impose restrictions on using global knowledge with a proper prompt query to retrieve only from the context. This approach of a RAG can reduce the hallucination to an extent.
\textit{Augmentation:} The augmentation step ensures the retrieved information is effectively passed on to the LLM context window. Augmentation can be anything that improves the performance of the LLM-generated response. 

Retrieval Augmented Generation is a potent method that enhances LLM's capabilities to produce reliable outcomes. In the RAG framework\cite{Lewis2020}, the output generation of LLM is enriched by integrating retrieved documents along with the query. This fusion of external knowledge sources aids in mitigating LLM's tendency to produce inaccurate or irrelevant information. Furthermore, it enables the assimilation of domain-specific knowledge, leading to improved outcomes without necessitating extensive fine-tuning. Additionally, employing RAG for content retrieval ensures the credibility of LLM-generated output and diminishes the need for extensive manual evaluation. These inherent features of RAG motivated our decision to apply it for both case law retrieval and structured summary generation. Moreover, RAG effectively handles lengthy content, rendering it highly suitable for applications in legal domains.

The proposed framework introduces an approach based on Retrieval Augmented Generation (RAG), specifically focusing on Legal Question-guided Retrieval by enhancing the query with facts and associated legal issues. We term this method Augmented Question-guided Retrieval (AQgR). This approach leverages an additional LLM module to generate legal questions based on the facts provided. When a user presents a question, the LLM internally employs mechanisms to understand and generate an appropriate response. Similarly, in the legal context, it first comprehends the legal facts and retrieves relevant judgments for the query. However, a single legal fact may entail multiple legal questions, and the LLM may not prioritize the most relevant ones. Hence, while the generated output may be accurate, it might not address the specific query sought by a legal professional. The proposed AQgR approach aims to resolve this issue.

The AQgR approach augments the RAG with a question of law or issue in the legal scenario to enable better case law retrieval. The existing approaches for case law completely rely on facts for precedent retrieval. In real-life scenarios, lawyers use legal questions to retrieve the precedents, not the facts. Here, the proposed method uses LLM to retrieve the precedents with legal questions and facts. A fact-based precedent retrieval system that identifies all possible legal issues in the underlying fact and provides relevant case laws with explanations can empower less experienced lawyers to make compelling arguments, even without extensive years of practice, by offering guidance and clarity in legal reasoning.
\begin{figure}[h]
\centering
\includegraphics[width=1.0\textwidth]{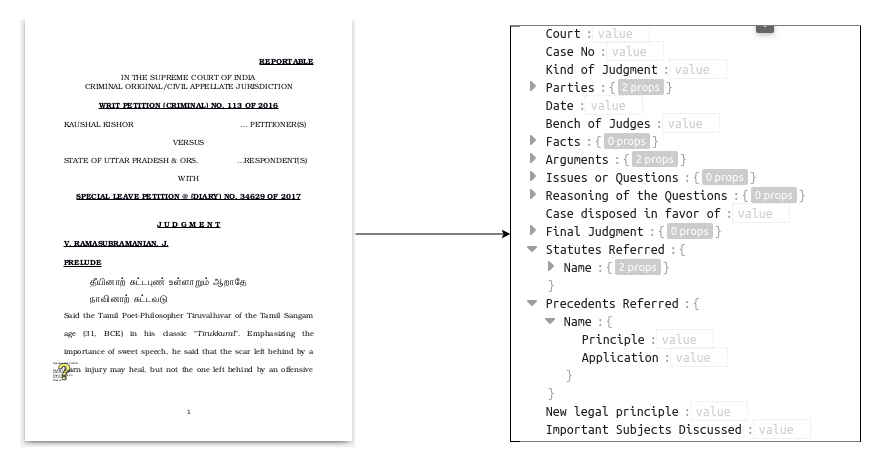}
\caption{The left side of the figure shows an unstructured judgment, and the right side shows the template created for building structured judgment for case law retrieval. }\label{fig2}
\end{figure}
\begin{figure}[h]
\centering
\includegraphics[width=1.0\textwidth]{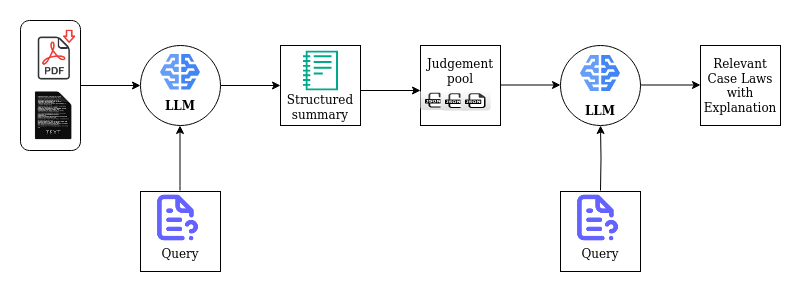}
\caption{Overview of the proposed framework using LLM}\label{fig7}
\end{figure}

\begin{figure}[h]
\centering
\includegraphics[width=1.0\textwidth]{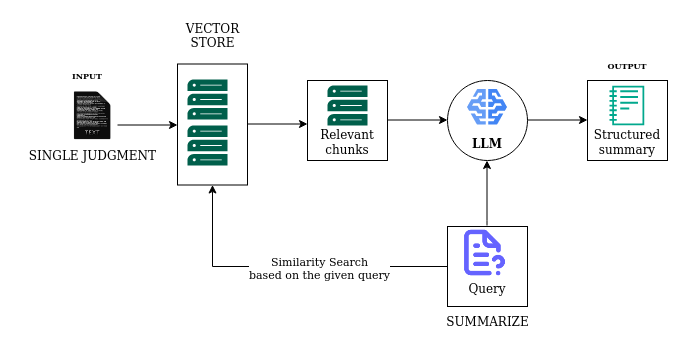}
\caption{Overview of the proposed framework for Structured summary generation using LLM}\label{fig3}
\end{figure}

\begin{figure}[h]
\centering
\includegraphics[width=1.0\textwidth]{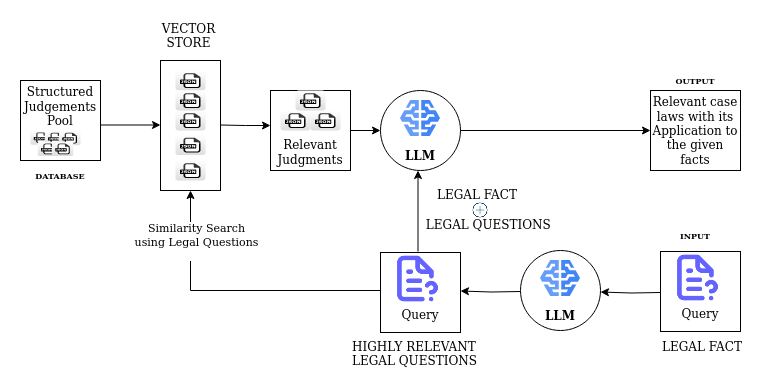}
\caption{Overview of proposed framework for case law retrieval using LLM}\label{fig4}
\end{figure}
The proposed framework \ref{fig7} follows a sequential process, commencing with generating structured summaries of available case laws. Developing structured summaries for training purposes with the aid of legal professionals can incur high costs. Hence, we employ the RAG-based approach to streamline the creation of these summaries more efficiently. Subsequently, a pool of judgments is compiled using the structured summaries. This judgment pool is an external knowledge source utilized alongside LLM for case law retrieval. The LLM query or facts of the judgment are provided as input to the LLM, which has access to the judgment pool with structured summaries. The LLM identifies relevant judgments based on the given fact and returns it in the order of relevance.

\subsubsection{Structured Summary Generation}

\begin{figure}[h]
\centering
\includegraphics[width=1.0\textwidth]{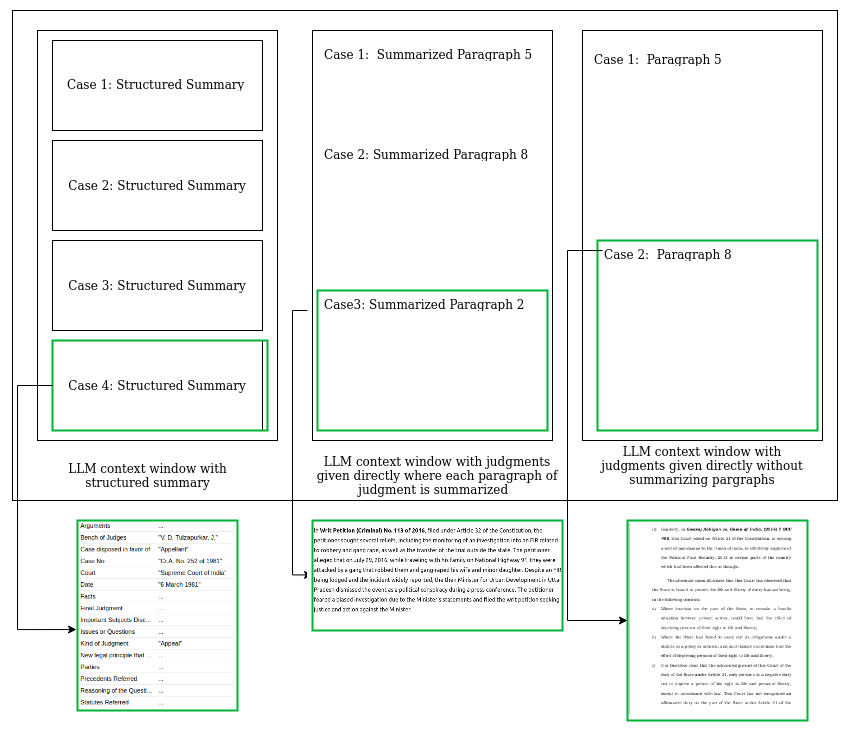}
\caption{LLM context window with structured summary of judgments, summarized paragraphs of judgments, judgments given directly without summarizing pargraphs }\label{fig5}
\end{figure}
The proposed framework endeavours to transform legal judgments into a structured format to manage document length effectively, thereby reducing usage costs and enabling the accommodation of more cases in the context window of LLM. Structured summary generation is not a new approach in case law retrieval. The authors in \cite{Ma2023} used a structured summary for case law retrieval, but they used a dataset of the Chinese law system where the documents are already structured. They just needed to summarize the content under each subtitle. However, for Indian Judgments, Rhetorical role segmentation(\cite{Noviello2023},\cite{Gupta2023},\cite{Bhattacharya2019rhetorical}) itself is a challenging task. For case law retrieval, a structured format minimizes the number of tokens and facilitates the organization and categorization of legal content.
\newline 
\newline

\textbf{Why do we need structured summary for case law retrieval?}

\begin{itemize}
    \item \textbf{Effective Document Management:} Structured summaries help manage lengthy legal judgments, reducing token usage and associated costs of processing documents in LLMs.

    \item \textbf{Enhanced Retrieval Efficiency:} By allowing for field-based searches, structured summaries improve retrieval by calculating similarity within specific fields rather than across the entire judgment, avoiding embedding loss.

    \item \textbf{Overcoming Unstructured Indian Judgments:} Unlike Chinese legal texts, which are often structured, Indian judgments require rhetorical role segmentation, making structured summaries essential for clear organization and categorization.

    \item \textbf{Broader Context Accommodation:} Structured summaries enable more judgments to fit within the context window of LLMs, expanding the pool of available case law for retrieval as depicted in figure \ref{fig5}. The figure shows three methods for accommodating legal judgments in a Large Language Model (LLM) context window for case law retrieval. Structured summaries, on the left, efficiently represent judgments by including key fields like case facts and reasoning, allowing multiple cases to fit within the context. Paragraph summarization, in the middle, condenses each judgment paragraph, fitting more information than full-text judgments but less efficiently than structured summaries. Direct text without summarization, on the right, uses full paragraphs, fitting fewer cases due to high token usage. Overall, structured summaries provide the most efficient approach for case retrieval. Elimination of unwanted fields in the structured summary based on application can further enhance the accomodation facility,

    \item \textbf{Legal Precision:} Summarized content must retain essential legal terminology to ensure accuracy, which structured summaries provide by extracting, rather than generating, content from judgments.

    \item \textbf{Cost-Effectiveness:} Structured summaries enable a cost-effective approach by allowing more judgments to be processed and retrieved without overwhelming LLM token limits.
\end{itemize}

Successful case law retrieval using the LLM necessitates the generation of summaries to accommodate more judgments in the context. However, in India's legal annotation context, generating structured summaries is both costly and challenging. Creating structured summaries for case law in India is expensive and difficult because legal judgments are often long, complex, and written in multiple languages. It takes legal experts to summarize these documents accurately, and their expertise comes at a high cost. The process is also tricky because legal texts can be interpreted in different ways, and ensuring high-quality, consistent summaries is time-consuming. There’s a huge number of cases to process, and current AI models have limits, like struggling with long texts or sometimes giving biased results. Additionally, setting up the technology to handle this and following rules to protect sensitive legal information adds to the cost and effort required.

To address this, an LLM is employed for structured summary generation without any fine tuning. In the context of case law retrieval, facts serve as input, with a straightforward approach involving finding the similarity score between the given fact and other judgments in the pool. Facts and the judgments pool are converted into embedding values to calculate similarity and retrieve relevant judgments for case law retrieval. However, while using LLM for structured summary generation, the facts generated using LLM will be mostly in normal English. When legal professionals use legal language for case law retrieval from the structured summary dataset which is in normal English lead to further issues or results in loss of important legal terms. So, to avoid that issue, an LLM is prompted to extract the relevant contents from the judgment and not generate the contents.

Additionally, legal professionals prefer extracted summaries for their reliance on specific details, particularly relevant paragraphs discussing legal principles. In Indian legal judgments, precedents are presented as paragraphs applicable to current case law. Considering this, the structured summary generation problem is modelled as a slot-filling problem by extracting relevant contents to fill the template. This method generates structured templates for both short and long judgments, facilitating the creation of a dataset for case law retrieval.

Structured summary generation plays a crucial role in facilitating efficient field-based searches. Employing a field-based search significantly enhances document retrieval efficiency. That is, instead of calculating the similarity of fact with the entire judgment, similarity can be calculated on a pool of facts. This field-based search enables better retrieval efficiency, avoiding embedding loss while calculating the similarity of fact with the entire judgment. A field-based retrieval based on specific relevant fields required for the application also helps accommodate more judgments in the LLM context window.

 The structured summary format as shown in figure \ref{fig2} includes judgment identification information, facts, questions or issues, reasoning, final judgment, statutes referred, and precedents .{The sample of a structured summary generated using proposed template is shown in Table \ref{tab:summary}.Each precedent is summarized with three key pieces of information: the name of the precedent, the reasoning principle used, and the reason for its application to the given facts. This comprehensive approach ensures that all relevant information about the precedents is also retained in the structured summary, thereby expanding the pool of judgments available for retrieval. For instance, if the LLM context window accommodates 30 structured summaries of judgments, the number of judgments available for retrieval would be 30. However, with complete precedent information in the structured summary, it will be much higher. This approach enriches the LLM context window for case law retrieval.

The selection of the LLM is done based on factors such as cost-effectiveness, reasoning capability, and token limit considerations from the leader-board\cite{llmleaderboard}. Given the high cost associated with newer models like GPT -4, more affordable yet effective options like Gemini Pro\cite{gemini} and GPT 3.5 Turbo\cite{openai} are chosen. The primary objective here is to extract structured summaries with less emphasis on the model's conversational abilities.

LLM models are integrated with RAG  to generate structured summaries for lengthy legal judgments. Despite the selection of two models, further experimentation is required to explore factors like chunk size, overlap, and vectors, aiming to gauge RAG's impact on processing long documents. Due to the high-cost implications of GPT models and our token limits, Gemini Pro is prioritized for further investigation due to its promising initial results. Additionally, incorporating chat history increases token limits, but chat history is essential for efficient Chain of Thought prompting. Utilizing GPT-3.5 models with chat history not only incurs significant costs but also decreases the size of the context window. The details of the chosen pipeline for structured summary generation using Gemini-pro model is given below.

\begin{itemize}
\item LLM used: Gemini pro \cite{team2023gemini}
\item Embedding: Google Generative AI embedding
\item Retriever: FAISS \cite{douze2024faiss}
\item Input token limit: 30720
\item Output token limit:2048
\item Type of learning: Zero-shot (without any examples in the prompt)
\item Query Type: Complex Query
\item Input: Single Judgment
\item Output: Structured summary of the given judgment
\item Prompt Template:
\begin{verbatim}
"""Context:\n{context}\n?
You:\n{question}\n
"""
\end{verbatim}
\item Prompt Query: " Summarize the given judgment by extracting contents such as Court: Case No: Kind of Judgment(Appeal/Petition): Parties: Date: Bench of Judges: Facts: Arguments: Issues or Questions: Reasoning of the Questions: Case disposed of in favour of: Final Judgment: Statutes Referred:{ Name: Principle: Application:} Precedents Referred:{ Name: Principle: Application:} New legal principle that can be applied to future cases:{ Principle: Application:} Important Subjects Discussed: Create structured output in JSON format. "
\end{itemize}

The structured summary generation process, illustrated in Figure \ref{fig3}, begins with receiving the judgment as input, which is then segmented into text chunks. These chunks are subsequently transformed into embeddings using Google Generative AI embedding, and the resulting vectors are stored in a FAISS vector store. Upon receiving the user's query, the embedded query is sent to the FAISS vector store to retrieve relevant chunks necessary to address the query through a FAISS similarity search. These retrieved chunks are integrated into an LLM context window alongside the query, enabling the generation of a structured summary. However, LLM tends to alter the structure for each run as it is a generative model. So, the output is created in a structured JSON format to have a common structure, at least for the parent nodes in the JSON file.

\subsubsection{Case Law Retrieval}

Case law retrieval involves identifying relevant legal precedents from a pool of judgments in response to specific legal queries. 	Automating case law retrieval has long been a goal in legal research. While existing methods can suggest relevant case laws, they often need more detailed explanations of why a particular case law is relevant to a given query. One of the key challenges in case law retrieval is the presence of diverse legal facts that may relate to the same legal issue. Traditional semantic similarity approaches struggle to identify different facts that address the same legal issue, especially within large pools of judgments. Our proposed approach addresses this challenge by explicitly linking legal issues with corresponding facts, thus facilitating more accurate retrieval.

While LLMs offer powerful contextual understanding capabilities, their effectiveness for case law retrieval can be limited by constraints such as context length and the need for indexing.

\textit{Context Length:} For instance, in a hypothetical scenario involving 50 judgments, each averaging around 20,000 words, the context length for LLMs like GPT-3.5 is limited to 4000 tokens. Since each token represents roughly 3-4 words in GPT-3.5, a 20,000-word judgment necessitates a context length of about 5000 tokens. During the search for relevant case law from a pool of judgments, matching chunks and their indices are retrieved based on a similarity score. If, for instance, four matching chunks are deemed relevant to the query, it is crucial that all four chunks collectively contain fewer than 4000 tokens for the LLM to process the context effectively. Ideally, these four chunks would originate from four different judgments, retrieving four distinct case laws as output. However, utilizing LLMs for case law retrieval presents a significant challenge due to the constraints of context length.

To address this issue, a practical solution involves condensing each judgment to retain only the most relevant information. A structured format typically comprises around 400 tokens, which allows the LLM to accommodate approximately ten condensed judgments within its context length. By adopting this structured format, we mitigate the challenge posed by LLM context length and facilitate more efficient information retrieval based on specific criteria.

\textit{Indexing:} The vector database stores each judgment chunk as a separate document. In a vector database, several whole judgments are saved as chunks with indexes as party names or any other judgment identification information. When a fact is given as a query, the retriever may identify some relevant paragraphs most similar to the given fact. The retriever retrieves the relevant chunks and passes them on to the LLM. However, LLM does not have access to the metadata of that chunk, and it cannot identify which judgment this chunk belongs to. Also, it can throw the available party names in that chunk as party names of the judgment if we ask for them in the prompt. This is one of the main practical challenges in case law retrieval using LLM.

However, the proposed approach solves all the above challenges using structured judgment. In the proposed methodology, each judgment is a different document, and each chunk involves the party names. So, the retrieved chunk involves the party names in the context itself. So, the LLM can give the output along with the party names or case no, whatever is needed.

Figure \ref{fig4} shows the proposed methodology involving a multi-step process, beginning with identifying legal issues corresponding to the given facts. These legal issues are then used to query a vector database containing structured judgments. The retrieved judgments are filtered based on relevance scores and passed through an ensemble retriever using FAISS\cite{douze2024faiss} and BM25\cite{bm25} algorithms. Finally, the selected judgments are analyzed by an LLM module to generate relevant details and assess their applicability to the given factual situation.

The details of the pipeline for case law retrieval are given below.

\begin{itemize}
\item LLM used: Gemini pro
\item Embedding: Google Generative AI embedding
\item Retriever: Ensemble(FAISS +BM25)
\item Input token limit: 30720
\item Output token limit:2048
\item Type of learning: Zero-shot and In-context learning
\item Query Type: Complex Query
\item Input: Multiple Judgments (Structured Summary)
\item Output: Relevant case laws with explanation and relevance score
\item Prompt Template:
\begin{verbatim}
generatelegalquestiontemplate ="""What are the three legal issues
or questions that are most relevant to the given facts:{question}?
Provide each relevant question as a list in descending order
of relevance to the fact.
{context}
Question: {question}
"""
retrievejudgmenttemplate ="""Just answer the Question: I will provide
you with some relevant parts of the prior judgments as context and facts
and its legal issues as the question. By reading the context, you must
provide me with the laws or judgments addressing similar legal
issues. Find relevant judgments(address more than one legal issue
given) applicable here with a relevance score(1-10) and the reason for
applying to the given facts. This list of judgments(Partynames)
should be sorted in descending order of relevance.

Context:\n {context}?\n
You:\n{question}\n
"""
\end{verbatim}
\item Prompt Query: Factual scenario
\end{itemize}

\section{Experiments \& Results}
This section presents a series of experiments to address the research questions outlined in the paper and evaluate the proposed methodologies. The experiments are designed to investigate the performance of LLMs in case law retrieval, assess the impact of structured summaries, examine the role of legal issues versus facts, explore the use of RAG, and evaluate the proficiency of RAG in generating structured summaries.
\subsubsection{Structured Summary Generation}
The experiments aimed to optimize various aspects of structured summary generation, including the selection of LLM, temperature settings, evaluation of the RAG framework, query formulation, prompt selection, chunking strategy, and retriever choice.

\textbf{ LLM Selection: }The LLM was chosen based on the token limit. The token limit is around 30,000 for the Gemini-pro model, which is very high compared to the GPT 3.5 model. Increased token limit models can accommodate lengthy judgments. The temperature and other parameters are set to default( temperature=0.9, top\_p=1.0, top\_k=1). Even though we have tried varying temperatures, the results were not good. While decreasing the temperature, LLM extracted the entire paragraph,  and the structured summary was almost the same size as the original judgment, especially in the case of small ones. So, the temperature value is reversed to default.

\textbf{Evaluation of RAG Framework}:  Five lengthy judgments were selected to evaluate the performance of the RAG framework. These judgments were obtained directly from courts in unstructured PDF format. Upon receiving a judgment, structured summary generation using RAG involves four steps.

\begin{enumerate}

    \item Split the judgments as chunks. The chunk\_size works based on the number of characters, and an overlap value is provided to ensure connectivity between chunks.
    \item After splitting the chunks, the chunks are converted into embedding and stored in a vector store.
    \item  Upon receiving the query embedding, the vector store finds similar embedding vectors and retrieves them from the vector store with the help of a retriever.
    \item The retrieved embedding vectors are stuffed onto the LLM context window along with the query.
\end{enumerate}

For each of these steps, different methods are used and chosen after many experiments. In the case of LLMs, the output is different for each run, so finding which factor contributes to a good output is challenging. 
\begin{enumerate} 
\item{ Parameter selection for Chunks}

\textbf{Chunking Strategy: }The recursive character splitter splits chunks based on paragraphs. As Indian judgments are structured as paragraphs, the recursive character splitter is used, and the results were good. The recursive character splitter splits based on the number of characters and paragraph splits in this scenario.\textit{ Chunk overlap value}: The chunk\-size and chunk\_overlap values, which are the arguments of recursive character splitter, are varied, and results are obtained. The number of chunks retrieved (the k value) and sent to the LLM is kept as default, equal to four. The chunk sizes chosen for the study include 1000, 5000, 10000, 20000 and 30000. Keeping chunk\_size as fixed, the overlap values are changed. The overlap values of  10\%, 30\%, 50\%, 70\%, and 90\%  of the chunk size are chosen. Keeping the chunk overlap value as 50\%  of the chunk\_size gave better results in terms of precision. Even though the template fields were filled with some value for 30\% chunk\_size, the quality was not good enough. Some fields were filled with wrong information. The fields were filled with null values for very high and low overlap values. For every chunk\_sizes, both smaller and larger, 50\% overlap value gave good results. \textit{Chunk size}: Finding a good chunk size is very important and tricky. Keeping the k value as 4, for large judgments, if the chunk\_size is very small, the retrieved chunks will be very small and will not contain much information, resulting in a bad output. Also, the number of tokens in all four chunks should be less than the context window size. Otherwise, it will not fit the context window and throw an error. Even if we decrease the value of \textit{k }(number of retrieved chunks) to 1 and increase the chunk size, essentially making it equal to the context window size, it would still result in output. However, it might overlook relevant information present in the other chunks. Therefore, maintaining the default value of \textit{k }  while reducing the chunk size proves to be a more effective approach. Considering the input token limit of 30,720, to ensure the four retrieved chunks utilize the entire context window, the size of each retrieved chunk should be 7500 tokens. The size of each retrieved chunk being 7500 tokens, with four retrieved chunks the entire context window will be filled with 30,000 tokens approximately. However, leaving some space within the context window is crucial to accommodate queries.
Moreover, overly large chunks may contain a small portion of relevant information amidst a large amount of irrelevant content, rendering them less useful and potentially leading to errors due to an overloaded context window size. Hence, setting the chunk size to 20,000 characters or 5000 tokens, with an overlap value of 10,000 characters or 2500 tokens are taken into account, and considering these factors yielded improved results. The four chunks will be there only for larger judgments. For smaller judgments,  the entire content fits within the first chunk itself leaving the other chunks empty.

\item{Selection of Embedding}

 Google Generative AI Embedding have been chosen to be compatible with the Gemini Pro model.

\item {Retriever Choice}

Two retrievers (Chroma and FAISS) are tested to select a suitable retriever. However, FAISS gave better retrieval quality and speed for this application. The parent document retriever has also been tested, and we hope for good results with this application. The parent document splitter initially splits documents into parent chunks in a parent document retriever. Then, each parent chunk is further split into child chunks to match the exact content during a similarity search. If the child chunk matches with a query, its parent chunk is retrieved and passed on to the LLM. However, there was no significant improvement in the results compared to  normal retriever.

\item {Query Formulation}

For summarization, the formulated query is complex enough to complete the structured summary generation task in one go for a single judgment. For a RAG-based LLM, a query is said to be complex if the query asks the LLM to give multiple information from multiple chunks. A query that asks for multiple pieces of information from the same chunk is a simple query for an LLM. However, the chosen query is complex and needs information from almost all the chunks. The chosen long judgments are downloaded court copies in PDF format. In India, information like court, party names, case number, kind of judgment, etc., are headers and will be on the document's first page. However, the Bench of Judges and Date will be at the end of the judgment. The facts, arguments, reasoning, ratio decidendi, and final judgment will be between the first and last pages. The query, as discussed in the earlier section for structured summary generation, is very complex as it asks the LLM to retrieve more than four relevant chunks from different parts of the document to answer the query. This is a problem in the case of longer documents with more than 1,22,880 characters. Here, three judgments are there with more than 1,22,880 characters. For those judgments, RAG was working, but not with complex queries. RAG-based LLM gave output for long judgments for complex queries, but the content precision and quality were poor. However, RAG worked well with simple queries. The Chain of Thought(CoT)\cite{Yu2023} prompting approach is also tested for long documents and works well for long documents.

\end{enumerate}

\subsubsection{Human Evaluation Results of LLM-Generated Structured Summaries}

 For case law retrieval, 50 judgments from the FIRE dataset are chosen, and structured summaries are created. As no dataset with structured summary is present for judgments present in the case law retrieval dataset, expert feedback is taken. A Google form is shared with legal research scholars and students. The quality of the structured summary generated for case law retrieval is evaluated. Each main field in the summary is evaluated with the help of metrics such as True positive, True negative, False positive and False Negative in terms of retrieval. An MCQ form asking about the retrieval quality of each field with TP, TN, FP, and FN is used for this purpose. To ensure fairness, a single judgment is evaluated by multiple experts. The results are depicted with a bar chart in figure \ref{fig6}. The quality of a structured summary is very important as it affects the quality of case law retrieval.

\begin{table}[h]
\caption{Details of the Judgments chosen for evaluating RAG}\label{tab0}%
\begin{tabular}{@{}llll@{}}
\toprule
Judgment & No. of Words  & No. of Characters \\
\midrule
CIVIL APPEAL NO.10941­10942 OF 2013   & 9968   & 56435  \\
CIVIL APPEAL NOS.1021-1026 OF 2013   & 18292  & 101107 \\
TRANSFERRED CASE (CIVIL) NO.98 OF 2012   & 37666  & 230848\\
WRIT PETITION (CRIMINAL) NO. 113 OF 2016& 70524& 417550\\
SLP (C) NOS.9036-9038 OF 2016)&108263&619812\\
\botrule
\end{tabular}
\end{table}

\begin{figure}[h]
\centering
\includegraphics[width=1.0\textwidth]{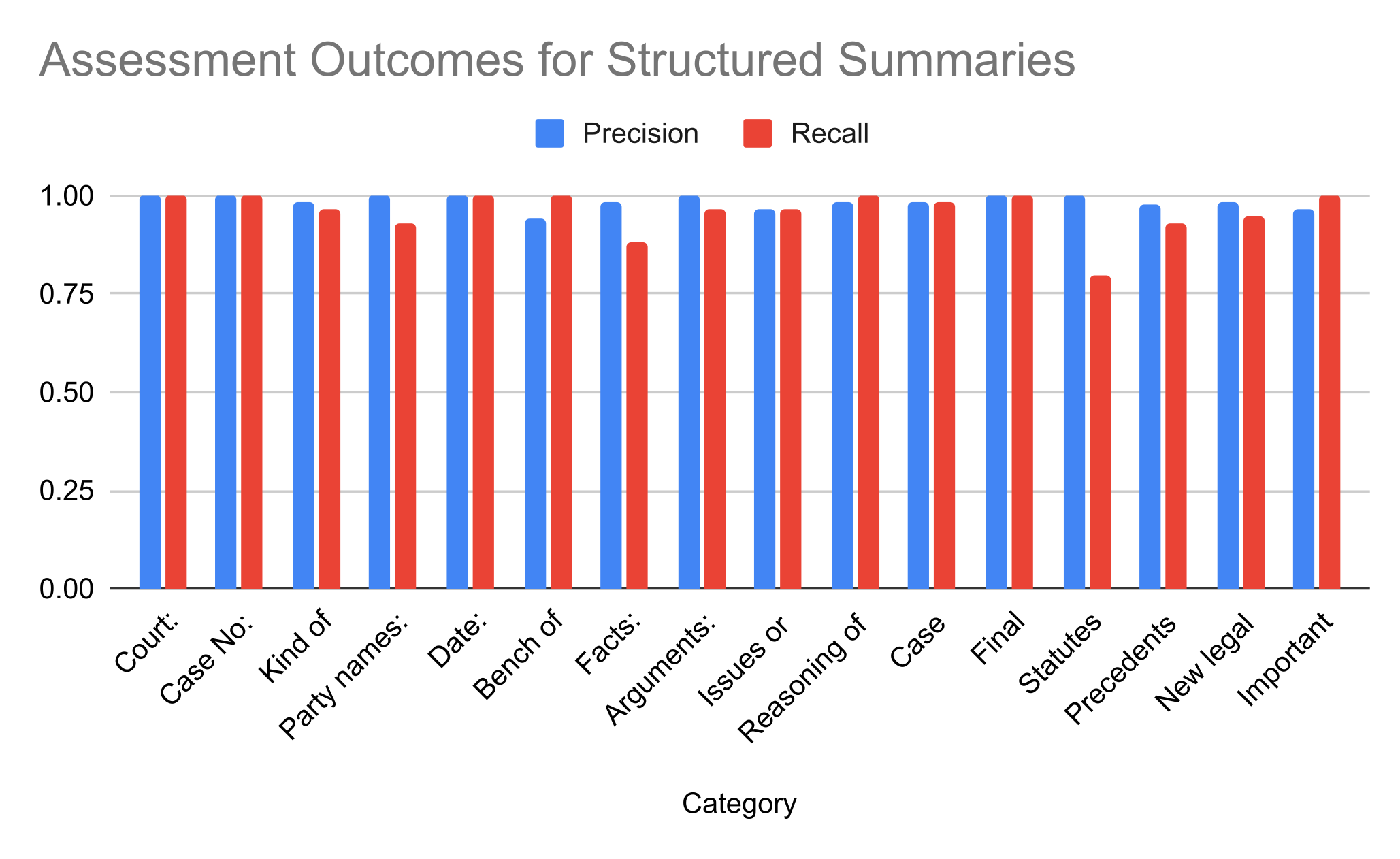}
\caption{Assessment Outcomes for Structured Summaries (Precision and Recall metrics for every category in the structured summary)}\label{fig6}
\end{figure}

\begin{table}[h]
\caption{Results of RAG evaluation on long documents for simple and complex queries}\label{tab3}%
\begin{tabular}{@{}llll@{}}
\toprule
Judgment & Simple & Complex & CoT \\
& Query&Query&prompting\\
\midrule

CIVIL APPEAL NO.10941­10942 OF 2013 & 5 &5&5\\
CIVIL APPEAL NOS.1021-1026 OF 2013 & 5&4&5\\
TRANSFERRED CASE (CIVIL) NO.98 OF 2012 &5 &2&5\\
WRIT PETITION (CRIMINAL) NO. 113 OF 2016 &5 &2&5\\
SLP (C) NOS.9036-9038 OF 2016&5 &2&5\\

\botrule
\end{tabular}
\end{table}

Figure \ref{fig6} illustrates the evaluation outcomes of the structured summaries created with LLM. Precision and recall values are computed for every main field extracted within the summary. The findings reveal a precision rate of nearly 90\% across most categories. However, the recall values for statutes, facts, and precedents are notably low. The reason for low values of recall, especially for statutes is that not every statutes are extracted by the LLM. A judgment may contain several statutes that are directly and indirectly related to the fact. However, the LLM extracted only relevant ones and not all the statutes in the judgment.

For evaluating the performance of RAG on long documents with simple and complex queries, five judgments in table \ref{tab0} is used. The summary generated for long documents does not use a standard query and is not evaluated based on metrics. The quality evaluation for long documents is done in different stages, with different queries handled by legal experts. After setting the parameters, a final test is done, and the experts are intended to give a rating (1-5) based on the quality of output for simple queries, complex queries and CoT prompting. The results are recorded in table \ref{tab3}. A RAG based LLM works for long documents but only with simple queries. For complex queries, CoT prompting gave better results.

\subsection{Case Law Retrieval}

For case law retrieval, several experiments were conducted to select the best settings that yield better results. Additionally, other experiments addressed research questions, such as the role of structured summaries, context length handling for practical implementation and AQgR in case law retrieval. Gemini Pro is utilized due to its high input token limit, making it suitable for case law retrieval. No chunking is needed in this application as each structured summary is separate, and there is no overlap between documents. Thus, each judgment is loaded and retrieved as a separate document. 

 An ensemble retriever comprising BM25 and FAISS is employed, as BM25 is a sparse retriever that excels in keyword-based searches. At the same time, FAISS is a dense retriever well-suited for semantic similarity matching. Combining both retrievers proves to be highly effective. Leveraging a portion of the FIRE dataset for case law retrieval provides ground truths, aiding in parameter tuning and prompt engineering to achieve the desired output. While numerous prompts were experimented with, the finalized prompt used for output retrieval is provided in the methodology section. 

Experiments were conducted with structured and unstructured judgments to evaluate the impact of structured summaries on case law retrieval. However, the retriever settings differ for each. Further chunking is unnecessary for structured judgments due to their small size and can be used directly. However, for whole judgments, both chunking and indexing are essential. A single judgment may be distributed across 3 or 4 chunks, requiring proper indexing to identify which judgment a specific chunk belongs to. Therefore, a parent document retriever is used. If a match is found in a smaller chunk, the entire chunk is retrieved and passed to the LLM. However, for the LLM, if that chunk lacks any indications of the details of the judgment, it may yield irrelevant answers or party names when queried for 'relevant judgments (party names) suitable for the given fact.' The whole judgment-based approach was ineffective, as it provided answers or explanations from the retrieved chunk of a specific judgment without considering the whole judgment, resulting in unsatisfactory results.

Using structured judgments for case law retrieval appears to be a favourable option as it retrieves entire judgment details as a single chunk, yielding remarkably better results. Despite querying the fact instead of legal issues, structured judgments provide more reliable explanations compared to whole judgments due to in-context learning. In structured summaries, facts and issues are paired with suitable precedents, enabling the LLM to better relate to the facts. However, while it provides explanations related to facts, the LLM identifies different legal issues or legal questions from the facts for each run, resulting in varying outputs.

To address this issue, an AQgR is enabled by adding an LLM module that generates legal issues or questions related to a given fact. The LLM generates numerous issues or questions of law, which are then sorted based on relevance, and three questions are selected for retrieving relevant chunks and generating LLM output. This method yields good results and reduces the randomness in generating different outputs for different runs. 

The next task is determining the number of structured cases accommodated within the context length. In real-life scenarios, the number of cases is expected to be high. Therefore, experiments were conducted to accommodate more cases by enabling the ensemble retriever to retrieve more cases. Typically, the retriever retrieves four judgments based on similarity search, but to determine the accommodation capacity, the retrievers are programmed to retrieve 12 relevant cases. Given an average length of 4000 characters for a structured summary, an LLM can accommodate 30 judgments. To prevent overflow, the limit of each retriever is set to 12, accommodating approximately 24 judgments. 

Further study aims to increase the system's capacity for practical implementation. By removing irrelevant fields and focusing solely on relevant information such as party names, judgment numbers, legal facts, questions, and reasoning, the average summary size is reduced to around 2000 characters, enabling Gemini Pro LLM to accommodate up to 60 judgments. The proposed AQgR system results are recorded  with the 'precedent referred' field and without the 'precedents referred' field. In the FIRE dataset, the results are purely based on judgments and not precedents contained in each judgment. So, results are recorded with and without a precedent field. 

\subsubsection{Results}
Two evaluation methods are used  for evaluating the quality of the output. Quantitative evaluation is done with Precision and Recall scores and Qualitative evaluation is done manually to analyze the quality of explanation generated. The Mean Average Precision(MAP) and Mean Average Recall(MAR) scores are utilized to evaluate the output of the case law retrieval. We tested the performance with 14 test queries and 50 Judgments. The query Q46 is omitted as it is throwing safety errors by LLM. For each query, the Average Precision (AP) of the retrieved documents is calculated. Average precision measures the average relevance of retrieved documents for a single query. However, ranking is not considered here, as that information was not present in the FIRE 2019 dataset. If the retrieved judgment is present in the list of relevant judgments, then it is considered for calculating the precision. The average recall is also calculated for each query for better understanding of the results.

The MAP is calculated by averaging the AP values across all queries and MAR is calculated by averaging AR across all queries. The results are recorded for the proposed method utilizing structured judgments with the 'precedents' field in the structured summary and without including the 'precedents' field. LLM generated a safety warning for some queries and judgments, so some judgments relevant to the query are excluded. For each query, the associated relevant judgments in the pool  are recorded in the table \ref{tab:case_law_results}, and precision and recall values are calculated based on that.

\begin{table}[h]
\caption{Precision and Recall for Case Law Retrieval with and without Precedent}\label{tab:case_law_results}
\begin{tabular*}{\textwidth}{@{\extracolsep{\fill}}llcccc}
\toprule
\multirow{2}{*}{Query} & \multirow{2}{*}{Judgment ID} & \multicolumn{2}{c}{Without Precedent} & \multicolumn{2}{c}{With Precedent} \\
\cmidrule{3-4} \cmidrule{5-6}
& & Precision & Recall & Precision & Recall \\
\midrule
Q1 & C14,C9 & 0.4 & 1 & 0.33 & 1 \\
Q2 & C27,C22 & 0.67 & 1 & 0.67 & 1 \\
Q3 & C1 & 0 & 0 & 0 & 0 \\
Q4 & C182 & 0.2 & 1 & 0 & 0 \\
Q5 & C54,C155,C121 & 0.33 & 0.33 & 0 & 0 \\
Q31 & C93,C65 & 0.67 & 1 & 0.67 & 1 \\
Q32 & C122,C164,C94 & 0.2 & 0.33 & 0.125 & 0.33 \\
Q33 & C186 & 0.33 & 1 & 0.25 & 1 \\
Q34 & C72,C49,C69,C25 & 0.25 & 0.5 & 0.29 & 0.5 \\
Q35 & C31,C184 & 0.2 & 0.5 & 0 & 0 \\
Q47 & C171 & 0.125 & 1 & 0 & 0 \\
Q48 & C82,C162,C141,C21 & 1 & 0.75 & 1 & 0.75 \\
Q49 & C174,C38,C76,C92 & 0.4 & 0.5 & 0.38 & 0.75 \\
Q50 & C27,C22 & 0.33 & 0.5 & 0.66 & 1 \\
\midrule
\textbf{Average} & & 0.3646 & 0.6721 & 0.3125 & 0.5236 \\
\bottomrule
\end{tabular*}
\end{table}

\begin{table}[ht]
\centering
\caption{Comparison with Baseline}
\begin{tabular}{|c|c|c|c|}
\hline
\textbf{Query} & \textbf{relevant GS Judgment ID}         & \textbf{AQgR(P/R)} & \textbf{Baseline (P/R)} \\ \hline
Q1             & C14, C9                     & 0.33 / 1.00                   & 0.10 / 0.50              \\ \hline
Q2             & C27, C22                    & 0.67 / 1.00                   & 0.00 / 0.00              \\ \hline
Q3             & C1                          & 0.00 / 0.00                   & 0.00 / 0.00              \\ \hline
Q4             & C182                        & 0.00 / 0.00                   & 0.00 / 0.00              \\ \hline
Q5             & C54, C155, C121             & 0.00 / 0.00                   & 0.20 / 0.67              \\ \hline
Q31            & C93, C65                    & 0.67 / 1.00                   & 0.10 / 0.50              \\ \hline
Q32            & C122, C164, C94             & 0.125 / 0.33                  & 0.10 / 0.33              \\ \hline
Q33            & C186                        & 0.25 / 1.00                   & 0.10 / 1.00              \\ \hline
Q34            & C72, C49, C69, C25          & 0.29 / 0.50                   & 0.20 / 0.50              \\ \hline
Q35            & C31, C184                   & 0.00 / 0.00                   & 0.10 / 0.50              \\ \hline
Q47            & C171                        & 0.00 / 0.00                   & 0.10 / 1.00              \\ \hline
Q48            & C82, C162, C141, C21        & 1.00 / 0.75                   & 0.40 / 0.75              \\ \hline
Q49            & C174, C38, C76, C92         & 0.38 / 0.75                   & 0.10 / 0.25              \\ \hline
Q50            & C27, C22                    & 0.66 / 1.00                   & 0.10 / 0.50              \\ \hline
\textbf{Average} &                            & \textbf{0.3125 / 0.5313}      & \textbf{0.1518 / 0.475}  \\ \hline
\end{tabular}
\label{tab:precision_recall}
\end{table}

The MAP and MAR values indicate that the quality of retrieval is good. Without including precedents in structured summary, our method scored an MAP of 0.36 and MAR of 0.67. With precedents, it scored an MAP of 0.31 and MAR of 0.52. Both exceeds the current state of the art MAP value which is 0.1573. A good MAR value indicates that the proposed system is able to retrieve ground truth relevant judgments to an extent in most of the queries.

For qualitative evaluation, the lawyers are given the results in a simple format.The queries in the FIRE 2019 dataset were lengthy, so with each output, the summary of query is also generated to make the evaluation easier. A sample of the case law retrieval output is included in the Appendix section. The feedback from the lawyers' community was really positive. They found the automatic generation of legal questions very helpful. Also, they appreciated the quality of the explanations provided.
\subsection{Comparison with Baseline}
The best performer in FIRE 2019 is chosen as Baseline\cite{Bhattacharya2019}. The baseline performance of the case law retrieval system is evaluated using 50 legal judgments and 14 test queries from the data set.  Each query is linked to a set of relevant judgments, serving as the ground truth. The baseline retrieves judgments based on textual similarity without considering precedent relationships like case citations. Judgments are ranked by their relevance score, using methods such as TF-IDF, BM25, or legal-specific embeddings.

The system's performance is measured using Precision and Recall. Precision reflects the proportion of retrieved judgments that are relevant, while recall indicates the proportion of relevant judgments successfully retrieved. Since the FIRE 2019 dataset lacks ranking information, any appearance of a relevant judgment in the retrieved list counts toward the calculation. Average Precision (AP) is computed for each query, and Mean Average Precision (MAP) is derived by averaging the AP scores across all queries. Similarly, Mean Average Recall (MAR) reflects the system’s overall recall.

This baseline serves as a benchmark to compare advanced models incorporating precedent knowledge, helping evaluate their effectiveness in improving case law retrieval.
\section{Result Analysis and Findings }

From the experiments and results, the following challenges are addressed.
\begin{itemize}
 
\item {Context Length Limitation of LLM:} The structured summary-based approach helps solve the context length limitations to an extent, and the structured summary with relevant fields helps solve the context length issue of LLM to a great extent and makes it suitable for practical applications. The structured summary also enabled in-context learning and helped  in relating facts to legal issues. For legal question generation, the fact which is given as input retrieves relevant contents from the database and given as context. The contents present in the context serves as examples and enables in\-context learning for question generation using LLM.

\item {Hallucination and cost of fine tuning:}The experiments show that RAG was very helpful in preventing the hallucination. With a good prompt query, RAG can reduce the hallucination to an extent. Even though RAG fails with complex queries for long judgments, it seems to be a promising solution with a CoT prompting even for long judgments. The RAG gave good results even without fine tuning for domain specific data. Also, RAG yielded better results with judgments that directly fits the context window of Gemini-pro and was able to create structured summary with just one query and zero shot learning.

\item{Retrieval Quality:} An ensemble retriever could give much better retrieval quality than a single retriever. The results returned by both retrievers were completely different but relevant ones. Some judgments that are not retrieved using the FAISS retriever for multiple runs are retrieved by the BM25 retriever and vice versa.

\item{Controlling the LLM Output:}The question-based retrieval gave good retrieval quality and helped maintain consistency in the LLM-based output to an extent, which enables the proposed LLM system for real-life implementation. However, the automatic legal issue generation module by LLM generates different legal questions for each run. The usage of the most relevant legal questions was able to solve the problem of generating different outputs for different runs. However, sometimes, legal questions generated by the LLM are more generalized and sometimes specific to the facts. A mechanism to control the LLM output while maintaining intelligence will be more helpful. Reducing temperature results in very low intelligence level leading to incorrect outputs. For practical application, the proposed approach should be modified in such a way that the lawyer can choose a legal issue from the generated list of possible legal issues. A fact combined with a single legal issue given as input can yield much better results. However,here we chose to select the three most relevant issues from the generated legal issues to test the potential of this approach with existing gold standard dataset.

\end{itemize}

The proposed method is tested both with and without references to previous cases, known as "precedents referred." When comparing against the gold standard results from the FIRE dataset, the method without "precedents referred" appears to be more effective because it doesn't take into account previous cases mentioned in the judgments. The decrease in precision and recall values for the method with "precedents referred" is because there are more relevant cases included in the pool (judgments and their precedents), and the structured summary contains more information. However, this abundance of information in the structured summary can lead to a poorer match with the query. Although the total number of retrieved judgments is higher when using "precedents referred," some of these judgments may only be indirectly relevant to the query in terms of statutes or facts. Nonetheless, legal experts believe that the method with precedents is more effective in practical scenarios. While the approach without "precedents referred" may include more relevant judgments in the context, the total number of relevant judgments will be lower. On the other hand, including "precedents referred" increases the total number of judgments in the context (judgments + precedents), and these precedents will be relevant to the given query as they are related to the facts in the actual judgment. Therefore, including precedents helps to include more relevant judgments in the context.

The findings of this research there by address the following questions.
\begin{itemize}
\item How do LLMs perform in case law retrieval withstanding context length restrictions and without fine-tuning?
The use of structured summaries and RAG helped in addressing the context length restrictions and the need for fine-tuning, respectively.

\item How do structured summaries contribute to case law retrieval using LLMs?
The use of structured summaries helped not only in accommodating more judgments but also in enabling in-context learning and generating better responses. The generated structured summaries have a lot of applications other than case law retrieval. Structured summaries were tested for statute law retrieval with reasoning for some test cases and yielded good results.
\item Contribution of legal issue in case law retrieval rather than facts.
The use of question-guided retrieval yielded better results than using facts. Additionally, legal experts found the generation of legal issues very useful.
\item Can Retrieval Augmented Generation (RAG) be used with LLMs to manage lengthy legal documents and diminish LLM hallucination?
RAG is a good solution for addressing hallucination issues. However, for generating structured summaries of long judgments, RAG did not perform well with complex queries. The CoT approach is a better solution for generating structured summaries of long judgments. Nonetheless, RAG remains an effective approach for case law retrieval with individual documents.

\item How proficient is RAG in generating structured or template-like summaries with just one query and zero-shot learning?
For small judgments that fit within the context window of LLM, a single complex query with zero-shot learning yielded good results. However, for long judgments, a single complex query with zero-shot learning did not work well.
\end{itemize}

\subsection{Limitations}
Despite the promising results, it is important to acknowledge the limitations of this study. The lack of full control over LLM outputs and the associated costs pose practical constraints on the widespread adoption of these models. Additionally, the evaluation of the proposed methodologies was conducted on specific datasets and scenarios, which may not fully represent the complexities encountered in real-world legal research.

Another limitation is the use of a single LLM model. Although we experimented with GPT, it was not a premium version, and our goal was not to identify the best LLM for this task. Instead, our focus was on assessing the general capabilities of LLMs for case law retrieval and structured summary generation in a cost-effective manner.

Furthermore, the approach’s reliance on structured summaries and external databases introduces dependencies that may impact scalability and generalizability. Precedent-based case law retrieval may also reflect biases or outdated legal principles, which can influence the reasoning process \cite{Ashley1992}. Addressing such outdated legal principles and biases was beyond the scope of this study.

However, our approach is adaptable for smaller-scale implementations, allowing legal professionals to use it with existing judgments, thus offering immediate benefits. This practical application could support lawyers' work as we continue to address broader challenges and limitations in the field.
\section{Conclusion}

In conclusion, this study’s experiments demonstrate the effectiveness of various methodologies for enhancing case law retrieval using LLMs. The structured summary approach, especially when augmented with relevant fields and precedents, proves promising in overcoming context-length limitations, making LLMs more viable for real-world applications. Additionally, the Retrieval Augmented Generation (RAG) framework effectively minimizes hallucination, enhancing the quality of LLM-generated summaries. Combining an ensemble retriever with a question-based retrieval approach further improves the relevance and consistency of retrieval results, providing legal professionals with more accurate and dependable information. These findings underscore the transformative potential of LLMs and advanced retrieval techniques for legal research and decision-making.

Automatically generating structured summaries of judgments via LLMs is a powerful technique with significant potential for addressing challenges in the Indian legal landscape. Structured summaries can facilitate statute retrieval, reasoning, argument generation, judgment prediction, and knowledge graph construction. Historically, the main drawback of Indian judgments has been their length and unstructured format, which posed obstacles to applying deep learning and transfer learning techniques, given their context-length limitations.

Future research can focus on developing more robust methods for controlling LLM outputs, exploring novel techniques to manage lengthy legal documents effectively, and enhancing the cost-efficiency of LLMs by addressing outdated legal principles and biases in case law retrieval. Additionally, there is a need for new evaluation mechanisms tailored to generative LLM outputs, as current metrics like ROUGE are insufficient for evaluating generative texts. New works using LLMs that involve generative text always require an expert in the field to evaluate the output, which is resource-intensive and costly. Expert evaluation is also feasible only for smaller subsets of the dataset.

\backmatter

\bmhead{Supplementary information}

The sample results of the AQgR for Case Law Retrieval is enclosed in the Appendix Section.

\begin{appendices}
The sample outputs generated by the proposed method is included in this section. The output for Query 1 and Query 2 is included. The change in the output patterns can also be seen from the sample. With same prompt, the output generated is in different format. The sample output generated by structured summary generation is also included.
\section{Sample Output}\label{secA1}
\subsection{Sample Output : Case Law Retrieval}
\textbf{Q1--------------------}\textit{In 1961, the appellant was appointed as an officer in a bank and later promoted to Foreign Exchange Department. In 1964, he became involved in a society formed for employee housing. In 1969, an inquiry found him negligent in handling society funds, resulting in a loss of Rs. 3,59,000. The bank suspended him in 1970 after criminal charges were filed against him. The inquiry officer later found him liable for Rs. 2,36,000 to the society. The bank terminated his services in 1971 based on this finding. The appellant successfully appealed the criminal conviction in 1973. He filed a writ petition in 1975 for reinstatement, which was granted by the High Court in 1979. However, the Division Bench of the High Court reversed this decision in 1985 on grounds of laches and merits.}
\begin{enumerate}
    
\item  Whether the Bank's termination of the appellant's services was legal and justified. \item  Whether the appellant was denied natural justice in the termination process. \item Whether the High Court erred in dismissing the appellant's writ petition on grounds of laches and merits.

\end{enumerate}
[
...
]
response
Relevant Judgments:
\begin{enumerate}
 
     \item  \textbf{Central Inland Water Transport Corporation Limited v. Brojo Nath Ganguly and Another}
        ◦ Relevance Score: 9 
        ◦ Legal Issue: Whether a service rule allowing termination of permanent employees without reasons is arbitrary and violative of Article 14 of the Constitution. 
        ◦ Reason: The judgment struck down a similar service rule as void and directed reinstatement with back pay. 
     \item  \textbf{West Bengal State Electricity Board and Others v. Desh Bandhu Ghosh and Others}
        ◦ Relevance Score: 8 
        ◦ Legal Issue: Whether a regulation allowing termination of permanent employee services without reasons is arbitrary and violative of Art. 14 of the Constitution. 
        ◦ Reason: The judgment held that such a regulation is arbitrary and violates Article 14. 
     \item  \textbf{O.P. Bhandari v. Indian Tourism Development Corporation Limited}
        ◦ Relevance Score: 7 
        ◦ Legal Issue: Whether a rule authorizing termination of an employee by giving notice or payment of salary in lieu thereof is constitutional. 
        ◦ Reason: The judgment held that such a rule is unconstitutional and violates Articles 14 and 16 (1) of the Constitution. 
     \item  \textbf{Ranchhodji Chaturji Thakore v. The Superintendent Engineer, Gujarat Electricity Board, Himmatnagar}
        ◦ Relevance Score: 6 
        ◦ Legal Issue: Whether an employee acquitted of criminal charges is entitled to reinstatement and back wages. 
        ◦ Reason: The judgment held that reinstatement should be granted but back wages could be denied for the period the employee was not in service due to conviction. 
     \item  \textbf{Jaipal Singh v. Union of India and Others}
        ◦ Relevance Score: 5 
        ◦ Legal Issue: Whether an employee acquitted of criminal charges is entitled to reinstatement and back wages. 
        ◦ Reason: The judgment held that reinstatement should be granted but back wages could be denied for the period the employee was not in service due to conviction. 
          
\end{enumerate}
\
\textbf{Q2------------------}\textit{The appellant, a former Cabinet Minister, testified as a witness in a bribery case against the first respondent, who was convicted. The High Court overturned the conviction, criticizing the appellant's conduct and credibility. The appellant had laid a trap for the respondent, who was arrested with marked currency in his briefcase. The prosecution claimed the respondent had demanded bribes monthly, while the respondent claimed the money was for donations. The High Court accepted the respondent's explanation and acquitted him, prompting the appellant to resign from his post to uphold democratic principles.}
\begin{enumerate}
   
     \item  Whether the disparaging remarks made by the Appellate Judge against the appellant, who was examined as prime witness in the trial and had become a Cabinet Minister at the time of the appeal, were justified and appropriate. 
     \item  Whether the High Court erred in acquitting the first respondent based on its finding that the prosecution had not proved its case by acceptable evidence and that the first respondent's explanation for the possession of the currency notes appeared probable. 
     \item  Whether the High Court's adverse remarks about the conduct of the appellant and the credibility of his testimony were supported by the evidence and whether they were fair and impartial. 
    
\end{enumerate}
[
...
]
response
\begin{enumerate}
  
\item  Whether the disparaging remarks made by the Appellate Judge against the appellant, who was examined as prime witness in the trial and had become a Cabinet Minister at the time of the appeal, were justified and appropriate.
Relevance Score: 10
    • Case Law:\textbf{ R.K. Lakshmanan v. A.K. Srinivasan and Another }
        ◦ Reasoning: The Court held that there was ample material before the High Court bearing on the impugned conduct of the appellant, justifying the adverse comments in question. 
\item  Whether the High Court erred in acquitting the first respondent based on its finding that the prosecution had not proved its case by acceptable evidence and that the first respondent's explanation for the possession of the currency notes appeared probable.
Relevance Score: 8
    • Case Law:\textbf{ Mohammad Naim v. State of Uttar Pradesh }
        ◦ Reasoning: The High Court has inherent power to expunge its own remarks to prevent abuse of process or secure justice. 
\item  Whether the High Court's adverse remarks about the conduct of the appellant and the credibility of his testimony were supported by the evidence and whether they were fair and impartial.
Relevance Score: 7
    • Case Law:\textbf{ K.R. Srinivas v. Dr R.M. Premchand }
        ◦ Reasoning: The Supreme Court held that Dr Premchand's writ petition was not maintainable as he had approached the High Court at a belated stage when the answer books had been destroyed, preventing effective cross-checking. 
  
\end{enumerate}
\subsection{Sample Output: Structured Summary Generation}

\title{Structured Summary of Judgment  \label{tab: summary}}

\begin{table}[h!]
    \centering
    \begin{tabular}{|>{\raggedright}p{3cm}|>{\raggedright}p{10cm}|}
        \hline
        \textbf{Key} & \textbf{Value} \\
        \hline
        \textbf{Court} & Supreme Court of India \\
        \hline
        \textbf{Case No} & Civil Appeal No. 562 of 1985 \\
        \hline
        \textbf{Kind of Judgment} & Appeal \\
        \hline
        \textbf{Parties} & 
        \begin{itemize}
            \item \textbf{Appellant}: West Bengal State Electricity Board and Others
            \item \textbf{Respondent}: Desh Bandhu Ghosh and Others
        \end{itemize} \\
        \hline
        \textbf{Date} & 26 February 1985 \\
        \hline
        \textbf{Bench of Judges} & O. Chinnappa Reddy, J. \\
        \hline
        \textbf{Facts} & Termination of respondent's services (a permanent employee) without reasons under Regulation 34 of the Board's regulations, allowing termination with three months' notice or salary in lieu. \\
        \hline
        \textbf{Arguments} & 
        \textbf{Appellant's Arguments}: 
        \begin{itemize}
            \item Regulation 34 does not offend Article 14 of the Constitution.
            \item Sections 18A and 19 of the Electricity Supply Act provide sufficient guidelines.
            \item Power to terminate services vested in higher-ranking officials, likely to be exercised reasonably.
        \end{itemize}
        \textbf{Respondent's Arguments}: 
        \begin{itemize}
            \item Regulation 34 is arbitrary and enables discrimination.
            \item The rule is a "hire and fire" policy, outdated and should be abolished.
        \end{itemize} \\
        \hline
        \textbf{Issues or Questions} & Whether Regulation 34 of the Board's regulations, allowing termination of permanent employee services without reasons, is arbitrary and violative of Article 14 of the Constitution. \\
        \hline
        \textbf{Reasoning} & 
        \begin{itemize}
            \item Regulation 34 is arbitrary and confers a power capable of vicious discrimination.
            \item It is a "hire and fire" rule with no guidelines or limitations.
            \item Similar rules have been struck down by this Court as violative of Article 14.
        \end{itemize} \\
        \hline
        \textbf{Case Disposed in Favor of} & Respondent \\
        \hline
        \textbf{Final Judgment} & Appeal dismissed with costs. \\
        \hline
        \textbf{Statutes Referred} & 
        \textbf{Electricity Supply Act}: 
        \begin{itemize}
            \item \textbf{Principle}: Guidelines for termination of services.
            \item \textbf{Application}: Sections 18A and 19 provide some guidelines, but not sufficient to save Regulation 34 from being arbitrary.
        \end{itemize} \\
        \hline
        \textbf{Precedents Referred} & 
        \textbf{Moti Ram Deka v. North East Frontier Railway}: 
        \begin{itemize}
            \item \textbf{Principle}: Rules allowing termination without inquiry or reasons may be contrary to Article 311(2) and Article 14 of the Constitution.
            \item \textbf{Application}: Cited as an example of a rule struck down for being arbitrary and violative of Article 14.
        \end{itemize}
        \textbf{S. S. Muley v. J.R.D. Tata and Ors.}:
        \begin{itemize}
            \item \textbf{Principle}: Standing Order allowing dismissal without inquiry or reasons is violative of natural justice.
            \item \textbf{Application}: Cited as an example of a rule struck down for being arbitrary and violating principles of natural justice.
        \end{itemize}
        \textbf{Workman, Hindustan Steel Ltd. v. Hindustan Steel Ltd.}:
        \begin{itemize}
            \item \textbf{Principle}: Standing Order allowing dismissal without inquiry or reasons violates natural justice.
            \item \textbf{Application}: Cited as an example of a rule struck down for being arbitrary and violating natural justice.
        \end{itemize}
        \textbf{Manohar P. Kharkhar v. Raghuraj}:
        \begin{itemize}
            \item \textbf{Principle}: Complexities of modern administration may necessitate powers like those under Regulation 48.
            \item \textbf{Application}: Cited by the appellant to support Regulation 34, but the Court disagrees with its reasoning.
        \end{itemize} \\
        \hline
        \textbf{New Legal Principles} & Arbitrary and uncanalised powers of termination of employment, without reasons or inquiry, are violative of Article 14 of the Constitution and natural justice; This principle can be applied to future cases involving challenges to termination rules or regulations. \\
        \hline
        \textbf{Important Subjects Discussed} & Arbitrary Power; Discrimination; Natural Justice; Termination of Employment. \\
        \hline
    \end{tabular}
    \caption{Structured Summary of Judgment}
    \label{tab:summary}
\end{table}

\section{Structured Summary of Judgments: Validation Questionnaire}
\author{}
\date{}

\maketitle

\noindent\textbf{Instructions:} \\
I will furnish you with a judgment along with its automatically generated structured summary. Your task will be to verify the precision of the structured summary. You'll need to validate each primary field of the summary. The original judgment (e.g., \texttt{C1.txt}) and its corresponding summary (\texttt{C1.json}) are located in a folder named \texttt{C1}. To examine the contents in a structured format, you can utilize the website \url{https://codebeautify.org/jsonviewer}. Simply paste all the contents from the JSON file provided to you on the website. This will present you with a structured format, facilitating easy validation.

\vspace{1em}
\noindent\textbf{* Indicates required question}

\vspace{1em}
\noindent\textbf{1. Email *} \\
\hrulefill
\vspace{1em}
\noindent\textbf{2. File name of the Judgments validated *} \\
\begin{enumerate}[label=\textbf{\arabic*.}, itemsep=0pt, parsep=0pt]
    \item C1.txt
    \item C9.txt
    \item C14.txt
    \item C21.txt
    \item C22.txt
    \item C25.txt
    \item C27.txt
    \item C31.txt
    \item C38.txt
    \item C47.txt
    \item C49.txt
    \item C54.txt
    \item C59.txt
    \item C65.txt
    \item C69.txt
    \item C72.txt
    \item C75.txt
    \item C76.txt
    \item C79.txt
    \item C82.txt
    \item C85.txt
    \item C92.txt
    \item C93.txt
    \item C94.txt
    \item C121.txt
    \item C122.txt
    \item C126.txt
    \item C139.txt
    \item C141.txt
    \item C147.txt
    \item C152.txt
    \item C155.txt
    \item C162.txt
    \item C164.txt
    \item C170.txt
    \item C171.txt
    \item C182.txt
    \item C184.txt
    \item C186.txt
    \item C2796.txt
    \item C2797.txt
    \item C2798.txt
    \item C2799.txt
    \item C2800.txt
    \item C2801.txt
    \item C2802.txt
    \item C2803.txt
    \item C2804.txt
    \item C2805.txt
    \item C2806.txt
\end{enumerate}

\vspace{1em}
\noindent\textbf{3. Name} \\
\hrulefill

\vspace{1em}
\noindent\textbf{4. Designation (Law student/Legal Professional) *} \\
\hrulefill

\vspace{1em}
\noindent\textbf{5. Court *} \\
\begin{itemize}
\item The system correctly identifies court name mentioned in the document and extracted it.
\item The system accurately determines that there's no court mentioned in the document.
\item The system incorrectly identifies court name. (Extracting wrong information)
\item The court name is present in the document but the system could not extract it.
\end{itemize}

\vspace{1em}
\noindent\textbf{6. Case No *} \\
\begin{itemize}
\item The system correctly identifies the case number mentioned in the document.
\item The system accurately determines that there's no case number mentioned in the document.
\item The system incorrectly identifies a case number as being present. (Extracting wrong information)
\item The case number is present in the document, but the system could not extract it.
\end{itemize}

\vspace{1em}
\noindent\textbf{7. Kind of Judgment (Appeal/Petition) *} \\
\begin{itemize}
\item The system correctly identifies the kind of judgment mentioned in the document.
\item The system accurately determines that there's no kind of judgment mentioned in the document.
\item The system incorrectly identifies a kind of judgment as being present. (Extracting wrong information)
\item The kind of judgment is present in the document, but the system could not extract it.
\end{itemize}

\vspace{1em}
\noindent\textbf{8. Party Names *} \\
\begin{itemize}
\item The system correctly identifies the parties mentioned in the document.
\item The system accurately determines that there are no parties mentioned in the document.
\item The system incorrectly identifies parties as being present. (Extracting wrong information)
\item The parties are mentioned in the document, but the system could not extract them.
\end{itemize}

\vspace{1em}
\noindent\textbf{9. Date *} \\
\begin{itemize}
\item The system correctly identifies the date mentioned in the document.
\item The system accurately determines that there's no date mentioned in the document.
\item The system incorrectly identifies a date as being present.
\item The date is present in the document, but the system could not extract it.
\end{itemize}


\vspace{1em}
\noindent\textbf{21. Overall Quality of the summary *} \\
\begin{itemize}
\item Very Satisfied
\item Satisfied
\item Neutral
\item Dissatisfied
\item Very Dissatisfied
\end{itemize}

\vspace{1em}
\noindent\textbf{22. Comments or Additional Features Needed in Future *} \\
\hrulefill




\end{appendices}

\bibliography{export}

\end{document}